# A Combination of Host Overloading Detection and Virtual Machine Selection in Cloud Server Consolidation based on Learning Method

Li Huixi[1,2], Xiao Yinhao[1,2] and Shen Yongluo[*1,2]
1. School of Information Science, Guangdong University of Finance and Economics, Guangzhou, 510320, China
2. Guangdong Intelligent Business Engineering Technology Research Center, Guangdong University of Finance and Economics, Guangzhou, 510320, China
*Correspondence: sylkyo@gdufe.edu.cn.

# Abstract

In cloud data center (CDC), reducing energy consumption while maintaining performance has always been a hot issue. In server consolidation, the traditional solution is to divide the problem into multiple small problems such as host overloading detection, virtual machine (VM) selection and VM placement and solve them step by step. However, the design of host overloading detection strategies and VM selection strategies cannot be directly linked to the ultimate goal of reducing energy consumption and ensuring performance. This paper proposes a learning-based VM selection strategy that selects appropriate VMs for migration without direct host overloading detection. Thereby reducing the generation of SLAV, ensuring the performance and reducing the energy consumption of CDC. Simulations driven by real VM workload traces show that our method outperforms the existing methods in reducing SLAV generation and CDC energy consumption.

# 1. Introduction

The Covid-19 outbreak has spurred an accelerated embrace of cloud computing over the past two years [1]. In order to maintain social distance and cut off the transmission of the virus, our work and life have moved online. Thanks to the help of cloud computing, the world economy under the shadow of the epidemic has remained basically normal [2]. On the other hand, this has also led to the development of cloud computing accelerating again.

The influx of various enterprises and individual users makes cloud service providers (CSPs) face severe challenges in allocating computing resources [4]. Cloud users hope to obtain satisfactory services, and cloud service providers also hope to convert computing power into services with minimal operating costs. Therefore, the CSP needs to consider the following complex relationship when configuring and scheduling various computing resources of the cloud data center (CDC): seeking a balance between maximization the user performance and minimization the energy consumption of the CDC [3]. Pursuing the tradeoff between energy consumption and performance has always been one of the key research issues in the field of CDC computing resource management.

## 1.1. Background

The quality of the service experienced by the users is a subjective feeling, but it can be measured by the computing power provided by the CSP. The relevant experience is quantified into a number of metrics described in the Service Level Agreement (SLA) for a direct assessment of service quality. SLA violation (SLAV) occurs when the metrics in the SLA cannot be met. Then the CSP should give the user a certain amount of compensation, which are factored into the cost calculation. On the other hand, energy consumption accounts for a large part of the cost of keeping a CDC running [5]. Generally speaking, the energy consumption of a CDC consists of two parts, which are the energy consumption of maintaining the operation of hardware devices (such as hosts) and the energy consumption of cooling equipment [6]. In this paper, we mainly consider the former. Because the CSP can directly reduce this part of energy consumption by optimizing task scheduling. For instance, a virtual machine (VM) packing algorithm is employed to save energy by packing virtualized devices (such as VMs and containers) into a smaller number of hosts, and then shutting down or switching additional empty hosts to power-saving modes [3]. In this server consolidation process, CSPs need to avoid a large-scale occurrence of SLAVs and a significant degradation of user experience caused by malicious competition for resources generated by excessively centralized VMs.

## 1.2. Motivation

In the study of server consolidation, a variety of over-loading detection strategies, VM selection strategies and VM placement strategies are proposed and implemented through combined execution [5]. The ultimate goal of server consolidation is to optimize the energy consumption or resource utilization efficiency of the CDC, but each strategy in the combination has a different purpose. When determining whether a host is overloading, there are strategies based on fixed thresholds and dynamic threshold strategies based on statistical methods. Due to the variety of criteria for determining whether a host is over-loading, when SLAV occurs, there must be excessive competition for resources among VMs, and when the over-loading state occurs, this competition does not necessarily occur at the same time. In other words, over-loading is not necessarily accompanied by SLAV, but there must be overloading when SLAV occurs. The purpose of filtering out over-loading hosts is to migrate some VMs in advance to prevent the generation of SLAV or to bring VMs out of SLAV state. Therefore, the goal of VM migration can be set as to avoid the occurrence of SLAV rather than the hosts overloading. When selecting VMs to be migrated, there are MMT that considers the migration time of VMs, random strategies, strategies that consider the minimum number of migrations, and MC strategy that considers the correlation between VMs and resources. The VM placement problem has always been a research hotspot in the field of cloud computing, and various algorithms have been proposed to solve this problem.

In solving the server consolidation problem, the above-mentioned combinations ignore the organic connection between various strategies. For instance, different CDCs, or even within the same CDC, need to deploy different VM placement algorithms to deal with different situations. In addition, the reason why overloading detection should be performed first in the server consolidation process is because the dynamic changes of VM workload make the emergence of SLAV uncertain. The host overloading threshold provides an early warning space for this uncertainty. In order to adapt to different placement algorithms and dynamic workloads, we use reinforcement learning to eliminate the host overload detection operation. In each time slot, our proposed method determines which hosts may be overloaded in the future, and also selects the VMs to be migrated, so as to achieve the goal of optimizing energy consumption and reducing SLAV.

# 2. Related Works

## 2.1. Non-machine-learning methods in Server Consolidation

### 2.1.1. Host over-loading detection strategies

Hieu et al. [12] proposed a host overloading detection strategy based on a multiple linear regression forecasting method. When the resource utilization of the host in the current and future specified time exceeds a preset fixed threshold, it is considered to be over-loading. YADAV et al. [8] proposed regression-based algorithms called Gdr and MCP to set a dynamic upper CPU utilization threshold for detecting overloading hosts. YADAV et al. [9] also proposed a over-loading host detection strategy based on robust regression, called LmsReg, which can dynamically adjust the upper CPU utilization threshold. Zhou et al. [10] proposed a KMI (K-Means clustering algorithm-Midrange-Interquartile range) strategy to distinguish underloading, normally loading and overloading hosts in a cluster. Fard et al. [11] proposed a host overloading detection strategy based on host CPU usage and temperature, both of which are fixed thresholds.

### 2.1.2. VM selection strategies

Hieu et al. [12] propose a VM selection strategy called minimum resource temperature (MRT), which selects those VMs that cause the most workloads to migrate. YADAV et al. [8] proposed Bandwidth-Aware Dynamic VM Selection Policy (Bw), which selects VMs with the smallest migration time to migrate. Yadav et al. [13] also proposed the MuMs VM selection strategy, which selects VMs with a smaller ratio of CPU usage to RAM usage on overloading hosts to migrate. Yadav et al. [11] proposed Minimum utilization prediction VM selection policy (MuP). This strategy uses statistical methods to analyze the CPU usage history of each VM to determine whether it should be migrated. Zhou et al. [10] divided the causes of host overloading into those caused by the use of CPU and those caused by the use of I/O. For the former, a strategy named MRCU (Maximum ratio of CPU utilization to memory utilization) is proposed. For the latter, a strategy named MPCU (Minimum the product of a CPU utilization and a memory utilization) is proposed. The maximum fit (MF) strategy was proposed by Fard et al [11]. The strategy first calculates the deviation between utilization of overloading server and its threshold, then picks the VM in which VM utilization is close to the deviation.

### 2.1.3. VM placement algorithms

Hieu et al. [12] uses PABFD. Zhou et al. [10] took SLAV into consideration and proposed a VM placement strategy with maximizing energy efficiency (VPME). Fard et al. [11] proposed a heuristic VM placement strategy to avoid SLAV. The core idea of this strategy is to classify hosts. A host with lower resource usage is less likely to experience SLAV in the future.

## 2.2. Reinforcement learning based methods in cloud

Rjoub et al. [14] proposed a reinforcement learning based method to improve the utilization of cloud resources and shorten the waiting time of tasks. Energy consumption and SLA are not considered. Qu et al. [15] proposed DMRO, which utilizes reinforcement learning to offloading tasks from IoT devices to edge-cloud servers. Considering the complex network structure between users and resources, Karthiban et al. [16] used reinforcement learning to achieve load balancing in the CDC. Sun et al. [17] proposed a reinforcement learning method to offload a computing mission from edge cloud to the vehicular cloud. Ma et al. [18] proposed a reinforcement learning-based model to evaluate which specific variables in the system are related to the energy consumption in a CDC. Joseph et al. [19] proposed to use RL methods to solve the problem of mapping a given set of Microservice containers to resources, aiming to reduce SLAV and energy consumption. Bitsakos et al. [20] proposed DERP, a reinforcement learning-based method that automatically scales various computing resources according to user needs, thereby achieving cloud elasticity. Cheng et al. [21] proposed DRL-cloud, which uses the RL method to distribute a large number of jobs to different servers, reducing energy consumption while ensuring that jobs can be completed before the deadline. In DRL-cloud, electricity price fluctuations are also set as constraints, so that energy costs are comprehensively considered. Tuli et al. [22] proposed a scheduling algorithm based on reinforcement learning and RNN to achieve seamless scheduling of application tasks between edge servers and cloud servers to reduce energy consumption and SLAV.

The above works did not focus on server consolidation, server load, and VM migration in CDC.

# 3. Problem Description and Proposed Methodology

In this section, we first describe the energy consumption model, then model VM migration and consolidation as a Markov Decision Process (MDP) and formulate the consolidation objective as the minimization of expected energy consumption increment.

## 3.1. Environment model

The life time of the cloud cluster system is divided into multiple time slots, which can be denoted as a sequence $[0, 1, 2, \cdots, t, t+1, \cdots, END]$. The length of time in a time slot is $T$.

There are $N$ VMs and $M$ hosts in the cluster. The $i$-th VMs is denoted as $vm_i$, and the $j$-th host is denoted as $host_j$.

$h_{j,t}$ is the allocation of VMs (or the set of running VMs) on the $host_j$ in $t$. $p_t = \{h_{1,t}, h_{2,t}, \cdots, h_{M,t}\}$ is the allocation of all VMs to the hosts.

The max amount of computing resource that $host_j$ can provide is its capacity $C_j$. To describe whether $host_j$ is in idle state or not in $t$, a binary indicator $\chi_{j,t} \in (0,1)$ is used. If $\chi_{j,t} = 1$, $host_j$ is not in idle state, and there is at least one VM running on it in time slot $t$. If $\chi_{j,t} = 0$, $host_j$ is in idle state, and it should be switched to energy-saving mode or shut down in time slot $t$.

## .3.2. Energy consumption model

Our energy consumption model is based on CPU utilization [5].

For $vm_i$, its energy consumption in $t$ is calculated as

$$ec_{vm_{i,t}} = \int_{t \cdot T}^{(t+1) \cdot T} u_{vm_i}(x) dx, \tag{1}$$

where $u_{vm_i}(x)$ is $vm_i$'s CPU utilization at time $x$.

For $host_j$, its energy consumption, $ec_{host_{j,t}}$, in $t$ consists of two parts. The first part, $base_{host_j}$, is the basic energy consumption when $host_j$ is running and in idle state. $base_{host_j}$ is constant. The second part is the energy that consumed by all the VMs running on it, which can be calculated as $\sum_{vm_i \in h_{j,t}} ec_{vm_{i,t}}$. Hence, we obtain the energy consumption of all hosts in time slot $t$:

$$EC_{host_t} = \sum_{j=1}^{M} \chi_{j,t} \times \left( base_{host_j} + \sum_{vm_i \in h_{j,t}} ec_{vm_{i,t}} \right). \tag{2}$$

It should be noted that $host_j$ dose not consume any energy if $\chi_{j,t} = 0$.

Migration cost in time slot $t$ should be considered. To migrate a VM, the cost is an extra 10% CPU usage of that VM at the source host. Hence, the total cost of VM migration is

$$MC_t = 10\% \times \sum_{vm_i \in MIG_t} ec_{vm_{i,t}}. \tag{3}$$

SLAV compensation of a VM in time slot $t$ is considered and is an important part of our optimization goal. If $host_j$ is experiencing 100% CPU utilization in $t$, it is overloaded and SLAV occurs. All VMs in $h_{j,t}$ should receive SLAV compensations at this time. A binary indicator $\Upsilon_{j,t}$ is used to tag whether $host_j$ is overloading and causes SLAV in $t$ or not:

$$\Upsilon_{j,t} = \begin{cases} 0, & if \sum_{vm_i \in h_{j,t}} ec_{vm_{i,t}} < C_j \\ 1, & if \sum_{vm_i \in h_{j,t}} ec_{vm_{i,t}} \geq C_j \end{cases}. \tag{4}$$

In this study, we map the SLAV compensation, which is denoted as $SLAVC_t$, to the energy consumption of related VMs:

$$SLAVC_t = c_{slav} \times \left( \sum_{j=1}^{M} \Upsilon_{j,t} \times \int_{t \cdot T}^{(t+1) \cdot T} \sum_{vm_i \in h_{j,t}} d_{vm_i} \right), \tag{5}$$

where $d_{vm_i}$ is $vm_i$'s initial CPU demand, and $c_{slav} \in [0, 1]$ is the SLAV penalty ratio. Since $d_{vm_i}$ is the user requirement and is a fixed value, we obtain

$$SLAVC_t = c_{slav} \times \left[ \sum_{j=1}^{M} \Upsilon_{j,t} \times \left( T \times \sum_{vm_i \in h_{j,t}} d_{vm_i} \right) \right]. \tag{6}$$

According to Equation 2, 3 and 6, we obtain total energy consumption $EC_t$ of the cluster in time $t$:

$$EC_t = EC_{host_t} + MC_t + SLAVC_t. \tag{7}$$

## .3.3. Markov Decision Process (MDP)

In the following, we model the MPD of our problem.

- State in time slot $t$: $s_t = p_t = \{h_{1,t}, h_{2,t}, \cdots, h_{j,t}, \cdots, h_{M,t}\}$.

- The action $a$: the migration actions over all VMS, which is combined by two steps.

  step 1: selecting VMs to migrate from each host. At time $t$, the set of selected VMs on $host_j$ is denoted as $MIG_{j,t}$. The set of all selected VMs is denoted as $MIG_t = \cup_{j=1}^{M} MIG_{j,t}$.

  step 2: implementing a given VM placement algorithm $P$.

  Hence $s_{t+1} = P(MIG_t, s_t)$.

  It should be noted here that our action is not a multi-level hierarchy model. Since our purpose is to train the decision-making policy in step 1, which can fit the given VM placement algorithm.

  The action space is denoted as $A(s)$.

- Reward $r(s, a, s')$: the direct reward of taking action $a$ at state $s$ and then arriving to certain new state $s'$.

- Policy $\pi(s)$: the strategy at state $s$, which is the probability distribution of actions at state $s$.

- Q-value $Q_\pi(s, a)$: the expected reward of taking action $a$ at state $s$ according to $\pi$.

## 3.4. VM selection and placement constraints

The action is combined by two steps, and in each step there are corresponding constraints. Regarding VM selection policy, it must obey the following constraint:

VM selection constraint: for each $host_j$, $MIG_{j,t} \subset h_{j,t}$, which means that the policy selects and only can select the VMs running on $h_{j,t}$ to be migrated.

Regarding VM placement policy, it must obey the following constraints:

1) VM placement constraint 1: every VM in $MIG_t$ must be assigned to a certain host;

2) VM placement constraint 2: the source host and the destination host of a migrating host

cannot be the same;
3) VM placement constraint 3: After the migration is complete, none of the VMs should be in the SALV state.

In this study, we use a modified PABFD algorithm [5] to allocate VMs to the appropriate hosts. The pseudo-code is presented in Algorithm 1. Comparing to the original PABFD, our algorithm modified the VM placement constraints and the core framework of the algorithm remains unchanged. Hence the modified algorithm is also called PABFD.

---

**Algorithm 1** PABFD

**Input:** hostlist, vmList( $MIG_t$ )

**Output:** allocation of the VMs
1:  **for** each VM *in* vmList **do**
2:      minPower ← MAX
3:      allocatedHost ← NULL
4:      **for** host *in* hostList **do**
5:          **if** *no SLAV on this host and not the source host for VM* **then**
6:              power ← estimatePower(host,VM)
7:              **if** power<minPower **then**
8:                  allocatedHost ← host
9:                  minPower ← power
10:             **end if**
11:         **end if**
12:     **end for**
13:     **if** allocatedHost != NULL **then**
14:         allocation.add(VM,allocatedHost)
15:     **end if**
16: **end for**
        **return** allocation

---

## 3.5. Energy minimization as migration goal

We define the reward function as the energy consumption decrement as action $a$ is taken at state $s$ and arriving at new state $s'$. It should be noted that the energy consumption is actually decreased at the new state if $r(s_t, a_t, s_{t+1}) > 0$. The goal is to design a VM selection strategy that minimizes the increment energy consumption:

$$r(s_t, a_t, s_{t+1}) = EC_t - EC_{t+1}. \tag{8}$$

The model is initialized as follows. $p_0$ is set to the VM allocation at time slot 0, and

$p_o = \{h_{1,0}, h_{2,0}, \cdots, h_{M,0}\}$. $Q_\pi(s,a)$ is 0, and $\pi(s)$ is uniformly distributed among all actions for each state. Then $Q_\pi(s_t, a_t)$ is updated by interacting with the cluster and server consolidation environment. The optimal strategy is given by the Bellman Equation such that the expected reward of taking action $a_t$ at state $s_t$ is the sum of $r(s_t, a_t, s_{t+1})$ and the future reward of $s_{s+1}$. Hence, we obtain

$$Q_\pi(s_t, a_t) = \mathbb{E}_{s_{t+1}}\left[r(s_t, a_t, s_{t+1}) + \gamma \mathbb{E}_{a_{t+1} \sim \pi(s_{t+1})}[Q_\pi(s_{t+1}, a_{t+1})]\right], \quad (9)$$

where $\gamma \in (0, 1)$ is the discounted factor of the future rewards.

The goal is to design a VM selection based migration strategy that maximizes the cumulative of energy consumption decrement value $r(s_t, a_t, s_{t+1})$ in the dynamic consolidation progress. We leverage the deep reinforcement learning method to solve the problem.

We use actor-critic based algorithm Proximal Policy Optimization (PPO) [24] to implement the VM selection based migration agent. We assume the probability ratio between old and new policies as

$$r_t(\theta) = \frac{\pi_{\theta_{new}}(a_t|s_t)}{\pi\theta_{old}(a_t|s_t)}. \quad (10)$$

The clipped surrogate objective function is:

$$J^{CLIP}(\theta) = \hat{\mathbb{E}}_t(\theta)\left[min\left(r_t(\theta)\hat{A}(s_t, a_t), clip(r_t(\theta), 1-\epsilon, 1+\epsilon)\hat{A}(s_t, a_t)\right)\right], \quad (11)$$

where $r_t(\theta)\hat{A}(s_t, a_t)$ is the normal policy gradient objective, and $\hat{A}(s_t, a_t)$ is the estimated advantage function. $clip(r_t(\theta), 1-\epsilon, 1+\epsilon)$ clips $r_t(\theta)$ to be in $[1-\epsilon, 1+\epsilon]$. The objective function of PPO is the minimum of the clipped and normal objective.

PPO restricts the range of large policy change as the incentive for the probability ratio to move outside of the interval is removed. Hence, the stability of the policy networks in improved in PPO by restricting the policy update at each training step. Another reason of leveraging PPO for VM selection based migration is that it is simpler to tune.

# 4. Performance Evaluation

## 4.1. Experimental Set Up

We use OpenAI [25] to simulate the CDC environment and the entire server consolidation process. PPO is one of the default reinforcement learning algorithms that comes with OpenAI [26].

Table 1. An example record of the Bitbrains VM workload trace

| TimeStamp[ms | CP | CPU capacity | CPU usage [%] | CPU | Memory | Memo | Disk | Disk | Networ | Networ |
|---|---|---|---|---|---|---|---|---|---|---|

| ] | U cores | provisioned [MHZ] | | usage [MHZ] | capacity provisioned [KB] | ry usage [KB] | read through put [KB/s] | write through put [KB/s] | k received through put [KB/s] | k transmitted through put [KB/s] |
|---|---|---|---|---|---|---|---|---|---|---|
| 1376322046 | 4 | 11703.99824 | 64.37199032 | 0.55 | 6.7108864E7 | 0.0 | 0.0 | 1.4 | 0.0 | 1.0 |

We use real workload trace dataset, Bitbrains [27], collected from VMs in production data centers to drive our simulations. By sampling every 5 minutes (a time stamp), Bitbrains recorded the usage of different resources by more than 4,000 VMs over a month. Table. 1 shows a sampling of resources used by a Bitbrains VM in a time stamp. In addition, Bitbrains also recorded the initial request for computing resources from these VMs.

We used Bitbrains trace to generate three sets of requests with different sizes. Request1 consists of 250 VMs, request2 consists of 500 VMs, and request3 consists of 100 VMs. We selected the resource usage records of related VMs within a random day (24 hours) to generate each request. Therefore, each request contains 288 time stamps, which can be used for 288 consecutive time slots in server consolidation. Based on these three requests, we measure the performance of the algorithm on different scale tasks.

## 4.2. Evaluation Metrics

Our goal is to reduce energy consumption and also reduce the generation of SLAV, and the corresponding indicators for evaluating the performance of the algorithm are EC and SLAV.

SLAV is calculated as follows. SLAV is related to 1) the SLAV time per active host (SLATAH): the percentage of time, during which active hosts have occupied the CPU utilization of 100% [5]; 2) the performance degradation due to the migration (PDM).

$$SLATAH = \frac{1}{M} \sum_{j=1}^{M} \frac{T_{s_j}}{T_{a_j}},$$

where $M$ is the number of hosts, $T_{s_j}$ is the SLAV time on $host_j$, and $T_{a_j}$ is the active time of $host_j$ [5].

$$PDM = \frac{1}{N} \sum_{i=1}^{N} \frac{C_{d_i}}{C_{r_i}},$$

where $N$ is the number of VMs, and $C_{d_i}$ is the estimation of performance degradation (can be estimated by an extra 10% of CPU utilization [23]) of the $vm_i$ caused by the migrations, $C_{r_i}$ is the total CPU capacity demanded by $vm_i$ [5].

$$SLAV = SLATAH \cdot PDM.$$

EC is calculated in 7. In addition, since VM migrations consume additional resources and degrade performance to some extent, the number of VM migrations is also used as one of the evaluation metrics.

## 4.3. Baseline Algorithms

The study by Buyya et al. [3,5] shows that the VM placement algorithm can show better results when the VM selection policy is MMT and the host overloading detection policy is LR. Therefore, in this paper, the combination of these two strategies is used to compare with our proposed method.
- MMT: the migration time is calculated by dividing the memory used by the VM by the available bandwidth. The MMT strategy selects the VM with the shortest migration time and adds it to the VM migration list;
- LR: the local regression method is used to count the history of CPU resource usage by the specified host, and then the strategy determines whether the host is overloading.

In addition, we also used two VM placement algorithms as the baselines, namely:
- Random placement (Random): for the current VM to be migrated, an available host is randomly selected in the host list as the destination;
- First Fit placement (FF): for the current VM to be migrated, FF selects the first available host in the host list as the destination.

We adopt LR-MMT-Random, LR-MMT-FF and LR-MMT-PABFD to compare with RL-PABFD, which is our proposed method, to evaluate its performance. The meaning of 'LR-MMT-Random' is to use the combination of LR, MMT and Random successively as the strategy for performing server consolidation.

## 4.4. Analysis of Results

## 4.4.1. EC

Figure. 1 demonstrates the total energy consumption after processing three sets of requests by various methods. It is shown in Figure. 1 that RL-PAABFD outperforms LR- MMT-Random, LR-MMT-FF and LR-MMT-PABFD regarding energy consumption. When the number of requested VMs is small, such as request1 and request2, the differences between the energy consumption generated by LR-MMT-PABFD and RL-PABFD are also small. When processing request3, RL-PABFD is significantly better than LR-MMT-PABFD. Under the same VM placement algorithm, RL does not show a sufficiently noticeable advantage over LR-MMT. The gap between RL-PABFD and LR-MMT-PABFD is far smaller than the gap between RL-PABFD and LR-MMT-Random. We believe that RL methods are not sufficiently convergent to deal with such problems. In addition, LR-MMT-Random, LR- MMT-FF and LR-MMT-PABFD all adopt the same VM selection policy and host overloading detection policy. Therefore, the disparity in energy consumption between the three is mainly caused by the performance of different VM placement algorithms.

Figure 2, 3 and 4 demonstrate the energy consumption in each time slot of the four methods when processing request1, request2 and request3. There are two main reasons for the fluctuation of energy consumption in different time slots: 1) dynamic changes in many MV workloads lead to changes in energy consumption, and 2) changes in the number of hosts used lead to an increase or decrease in energy consumption. Among them, the drastic changes in the workload of the VMs are the main reason for this fluctuation. Therefore, when processing a given request, although the energy consumptions of the CDC generated by the four methods are constantly changing, some are more and some are less, but in most time slots, their overall trends are relatively close. For instance, in certain time periods, if the resource demands for resources of a large number of VMs increases significantly, the energy consumption of VMs will also inevitably increase. In order to reduce SLAV, more hosts must be used to load these VMs, thereby energy consumption is further increased. This inevitable increase in energy consumption cannot be substantially reduced by algorithms. The underlying algorithms cannot influence the resource demands of higher-level users or tasks.

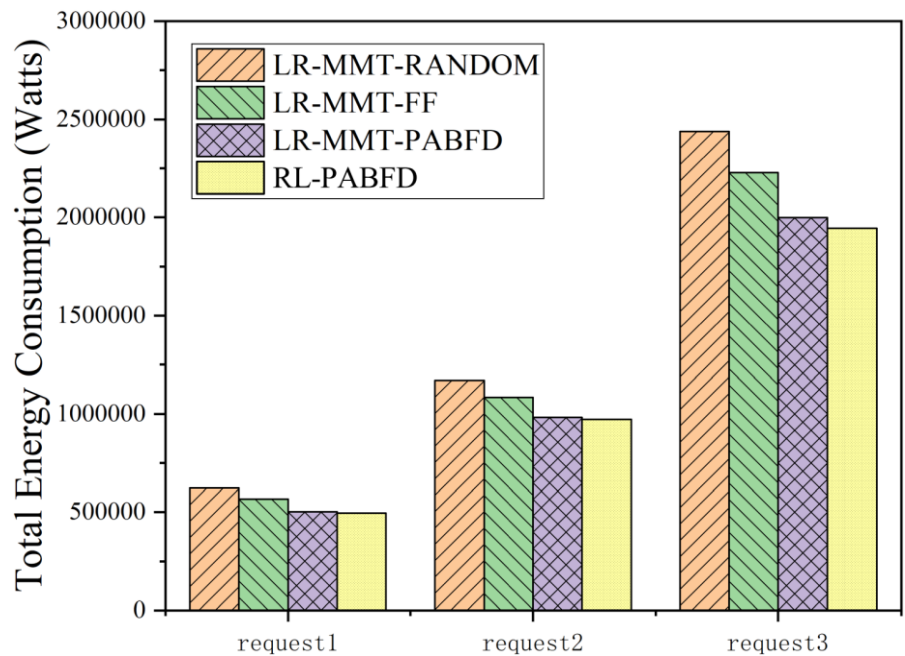

Figure 1. Comparing the total EC of four methods regarding request1, 2 and 3.

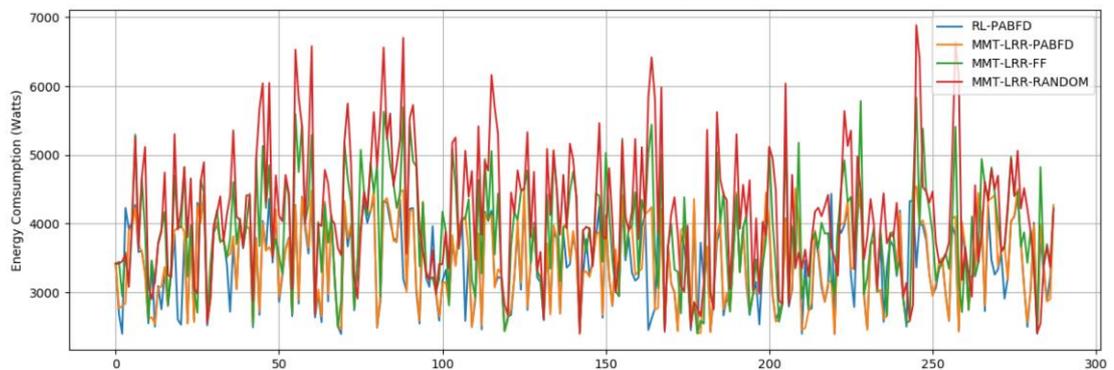

Figure 2. Comparing EC in every time slot of four methods regarding request1.

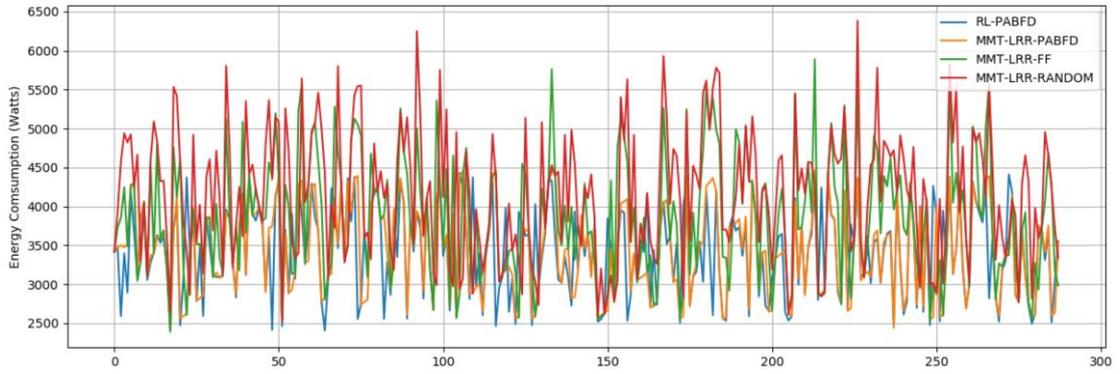

Figure 3. Comparing EC in every time slot of four methods regarding request2.

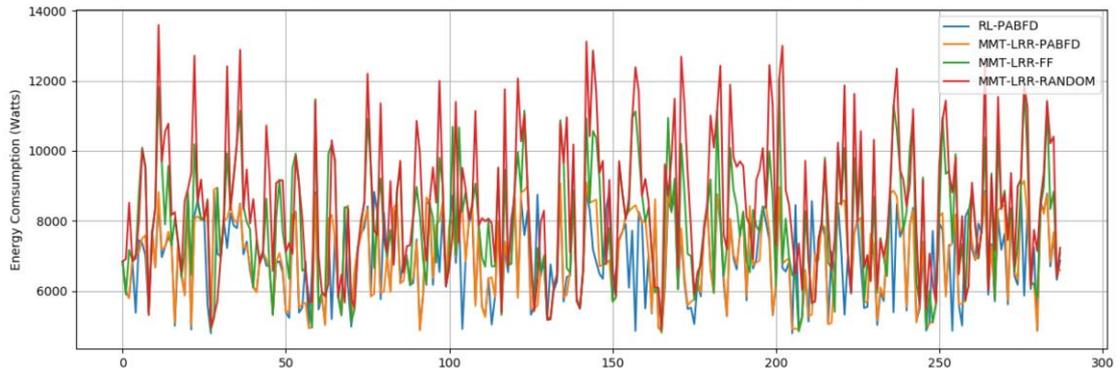

Figure 4. Comparing EC in every time slot of four methods regarding request3.

## 4.4.2. SLAV

Figure 5 demonstrate the SLAV generated by various methods after processing three sets of requests. It can be seen from Figure 5 that when processing requests of various sizes, the SLAV generated by RL-PABFD is significantly less than that of other methods. This is because part of the objective function, as shown in 8, consists of the SLAV penalty when we design the solution. Therefore, the policy will actively reduce the generation of SLAV during learning, thereby avoiding the increase of energy consumption. LR-MMT can only passively and indirectly reduce the possibility of SLAV generation, and their purpose is not directly reflected in the final optimization goal. As can be seen from Figure 5, comparing LR-MMT-Random, LR-MMT-FF and LR-MMT-PABFD, RL-PABFD is more capable of reducing SLAV than reducing overall energy consumption.

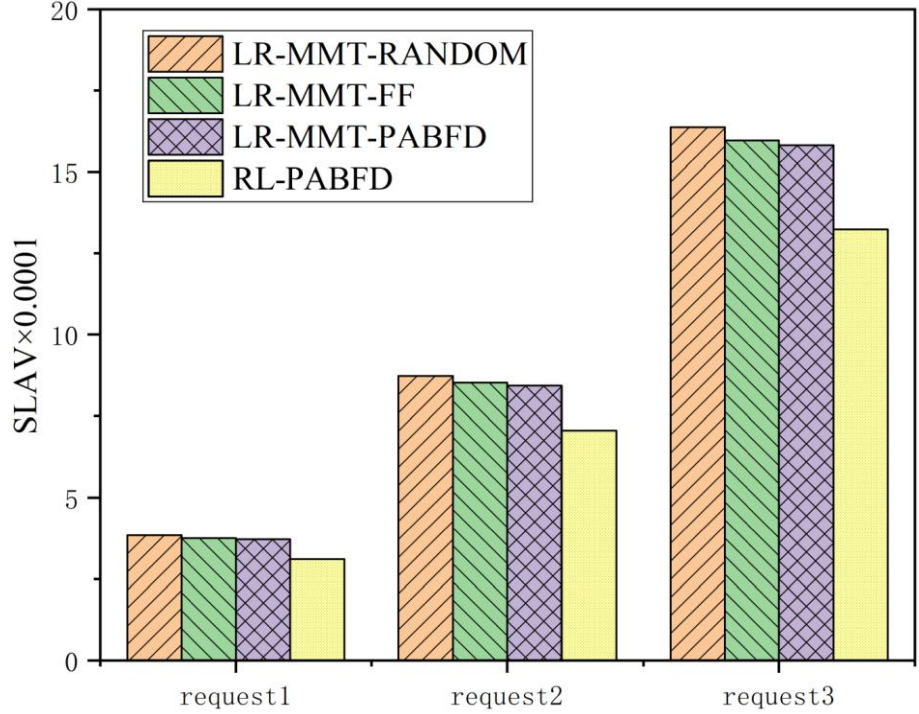

Figure 5. Comparing SLAV of four methods regarding request1, 2 and 3.

Figure 5 also shows that the SLAV gaps among LR-MMT-Random, LR-MMT-FF and LR-MMT-PABFD are smaller than that among RL-PABFD and them. This is because the three use the same combination of strategies, LR-MMT, to prevent the host from entering the overloading state or staying in this state for a long time. The generation of SLAV is mainly related to the host overloading detection strategy and VM selection strategy. The primary responsible goal of the VM placement algorithm is to select the appropriate host for each VM to be migrated. The constraints or optimization objectives related to SLAV are not considered by the Random, FF, and PABFD algorithms.

During the server consolidation process, multiple VMs were in the SLAV state in multiple consecutive time slots, so we will not show and compare the SLAV generated by the four methods in every time slots here.

## 4.4.3. Number of VM migrations

Figure 6 shows the total number of VM migrations trigged by various methods after processing three sets of requests. As can be seen from Figure 6, RL-PABFD results in the smallest number of total VM migrations when processing requests of various sizes. However, compared with the other three methods, RL-PABFD does not show an absolute advantage. Although VM migrations will generate additional energy consumption, its purpose is the load balancing of the cluster, thereby better reducing SLAV. There is a tradeoff between the two, but the focus is on reducing SLAV. Comparing Figure 6 with Figure 5, it can be seen that RL-PABFD reduces SLAV more than the other three methods when the difference in the total number of virtual machine migrations induced is small. This also means that, compared with the other three methods, RL-PABFD can select a more suitable VMs for migration, which can lead to SLAV generation when resources are fiercely contested. Observed from this perspective, RL is more efficient than LR-MMT.

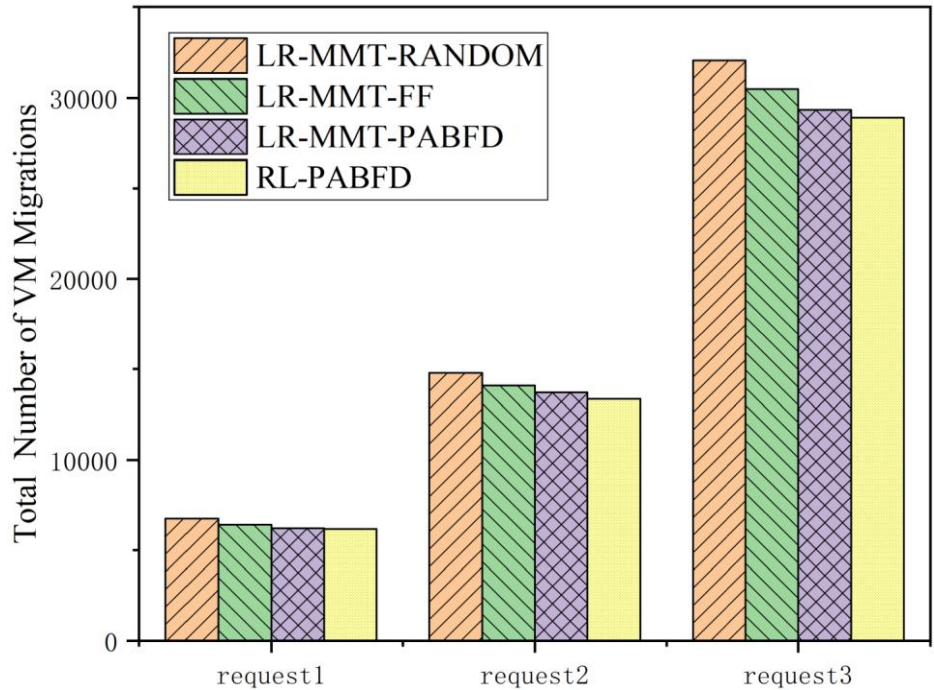

Figure 6. Comparing the number of VM migrations of four methods regarding request1, 2 and 3.

Figure 7, 8 and 9 demonstrate the number of VM migrations trigged by the four methods in each time slot when processing the three requests. From these figures, it can be seen that the number of VM migrations induced by the four methods tends to the same trend in multiple consecutive time slots. However, within a single time slot, the fluctuations in the number of virtual machine migrations caused by the four methods are different. This phenomenon is similar to the EC fluctuation phenomenon shown in Figure 2, 3 and 4. The reason for this phenomenon is still that the fluctuation of a large number of VM workloads leads to fluctuations in the demand for computing resources. When most of the VMs are about to enter or are already in the peak business period, the centralized demand for computing resources will trigger the generation of SLAV and a large amount of energy consumption. Unlike reducing energy consumption, SLAV can only be eliminated by VM migrations.

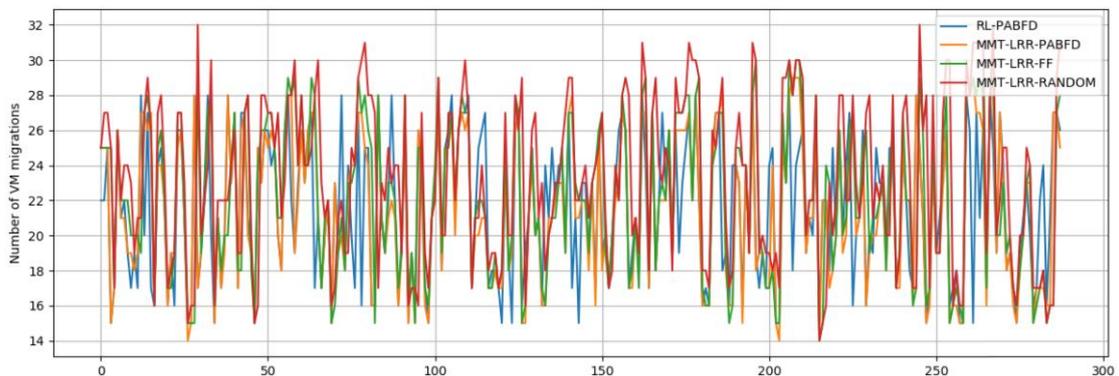

Figure 7. Comparing the number of VM migrations in every time slot of four methods regarding request1.

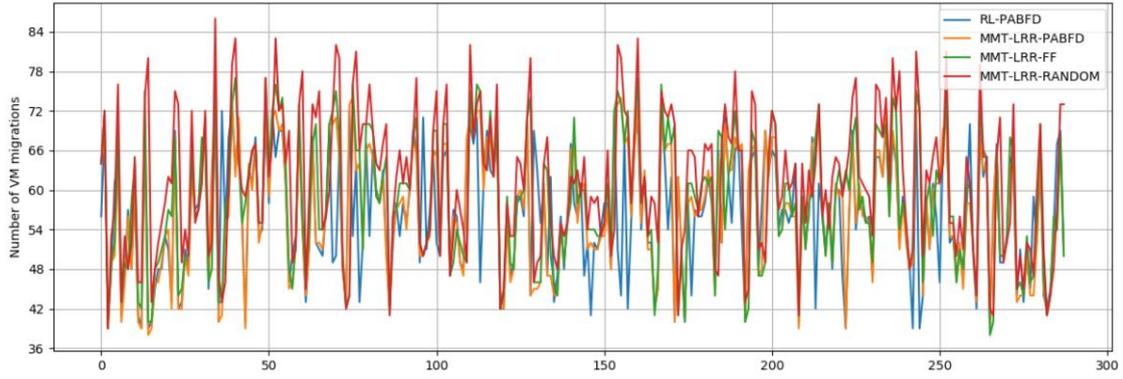

Figure 8. Comparing the number of VM migrations in every time slot of four methods regarding request2.

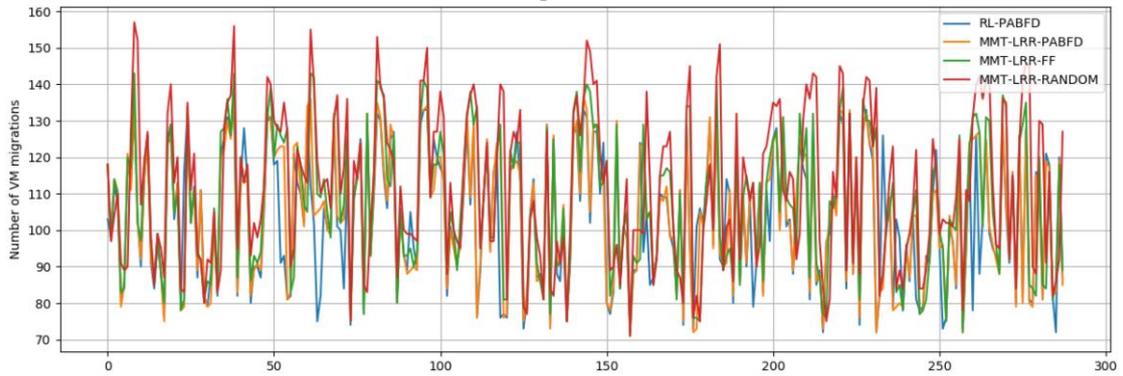

Figure 9. Comparing the number of VM migrations in every time slot of four methods regarding request3.

# 5. Conclusions

This study proposes a reinforcement learning based VM selection strategy, which selects appropriate VMs for migration from hosts that may be in or have been overloaded, so as to avoid the occurrence of SLAV and reduce the total energy consumption of CDC. Our proposed method combines the two steps of host overloading detection and VM selection during server consolidation, and unify the two with the ultimate goal of reducing energy consumption. Simulations driven by real VM workload traces demonstrate that our method is slightly better than the existing methods in reducing the energy consumption of SLAV generation of CDC, showing a certain feasibility.

In future work, we will further improve the convergence performance of the proposed method and also incorporate VM placement into the consideration of using RL-based ideas to solve the server consolidation problem holistically.

# Acknowledgement

This research was funded by National Natural Science Foundation of China (No.62002067), Guangzhou Youth Talent Program (QT20220101174), Department of Education of Guangdong